

\def\f{\phi}
\def\l{\lambda}\def\m{\mu}\def\n{\nu}\def
\p{\pi}\def\s{\sigma}\def\t{\tau}

\def\mo{{-1}}\def\ha{{1\over 2}}

\def\gmn{g_{\m\n}}\def\ghmn{\hat g_{\m\n}}

\def\dx{\int d^2x\ \sqrt{-g}\ }\def\ds{ds^2=}
\def\dhx{\int d^2x\ \sqrt{-\hat g}\ }

\def\st{spacetime }
\def\bh{black hole }
\def\tran{transformations }\def\coo{coordinates }
\def\bg{background }\def\gs{ground state }
\def\sm{semiclassical }\def\hr{Hawking radiation
}\def\sing{singularity }

\def\sch{Schwarzschild }\def\min{Minkowski }\def\ads{anti-de Sitter }

\def\section#1{\bigskip\noindent{\bf#1}\smallskip}

\def\ef{e^{-2\f}}\def\ely{e^{\l(y^+-y^-)}}
\magnification=1200

\font\titolino=cmbx10
\font\tsnorm=cmr10
\font\tscors=cmti10

\font\tscorsp=cmti9
\magnification=1200

\def\PRD{{\tscors Phys. Rev. D }}

\def\NPB{{\tscors Nucl. Phys. B }}

\def\PLB{{\tscors Phys. Lett. B }}

\def\note{\advance\notenumber by 1 \footnote{$^{\the\notenumber}$}}
\def\ref#1{\medskip\everypar={\hangindent 2\parindent}#1}
\def\beginref{\begingroup
\bigskip
\leftline{\titolino References.}
\nobreak\noindent}
\def\endref{\par\endgroup}
\def\beginsection #1. #2.
{\bigskip
\leftline{\titolino #1. #2.}
\nobreak\noindent}

\nopagenumbers
\null
\vskip 5truemm
\rightline {INFNCA-TH9516}

\rightline { }
\rightline { }
\rightline{May 1995}
\vskip 15truemm
\centerline{\titolino ON THE CONFORMAL EQUIVALENCE BETWEEN}
\bigskip
\centerline{\titolino 2D BLACK HOLES AND RINDLER SPACETIME}
\vskip 15truemm
\centerline{\tsnorm Mariano Cadoni and Salvatore Mignemi}
\bigskip
\centerline{\tscorsp Dipartimento di Scienze Fisiche,}
\smallskip
\centerline{\tscorsp Universit\`a  di Cagliari,}
\smallskip
\centerline{\tscorsp Via Ospedale 72, I-09100 Cagliari, Italy.}
\smallskip
\centerline {\tscorsp and}
\smallskip
\centerline{\tscorsp INFN, Sezione di Cagliari.}
\bigskip
\vskip 15truemm
\centerline{\tsnorm ABSTRACT}
\begingroup\tsnorm\noindent
 We study a two-dimensional dilaton gravity model related by a
 conformal transformation of the metric to the
 Callan-Giddings-Harvey-Strominger model. We find
 that most of the features and problems of the latter can be
 simply understood in terms of the classical and semiclassical
 dynamics of accelerated observers in two-dimensional
 Minkowski space.


\vfill
\leftline{\tsnorm PACS:04.70.Dy,11.25.-W, 97.60.Lf\hfill}
\smallskip
\hrule
\noindent
\leftline{E-Mail: CADONI@CA.INFN.IT\hfill}
\leftline{E-Mail: MIGNEMI@CA.INFN.IT\hfill}
\endgroup
\vfill
\eject
\footline{\hfill\folio\hfill}
\pageno=1

In the past few years, a great deal of work has been dedicated to the
study
of a two-dimensional model of dilaton gravity first proposed by
Callan et al. (CGHS) [1].
The reason of this interest is due to the fact that such model
provides a
good approximation of low-energy scattering from a nearly extremal
\bh of four-dimensional string gravity.
Such approximation is  much more tractable than the
original model and in fact permits the investigation of  the process
of
formation and evaporation of a \bh in a semiclassical approximation.

The model is described by the two-dimensional action:
$$S=\dhx\left[\ef(\hat R+4(\hat\nabla\f)^2+4\l^2)-\ha\sum(\hat\nabla
f_i)^2\right],\eqno(1)$$
where $\f$ is the dilaton and $f_i$ are a set of scalar fields.

At the classical level, this model admits an exact solution
describing
the
formation of a \bh caused by a shock wave of incident matter. It is
then
possible to discuss the Hawking radiation of the \bh by means of the
standard
\sm calculation performed by quantizing the matter fields $f_i$ in
the
\bg
constituted by the classical solution. As is well known, in this
approximation
the contribution of the scalars to the energy-momentum tensor is
proportional
to the one-loop anomaly, which in the conformal gauge is in turn
proportional to
the curvature of the \bg metric.

The final result of the calculation is that a constant flux of
radiation is
emitted, which is independent of the mass of the \bh [1]. This
surprising
result can be improved by making a better approximation, taking into
account
the backreaction of the gravitational field to the radiation. This
topic has
been widely investigated [2].

A further problem arises when one considers conformal \tran of the
original
metric. As is well known, in fact, a conformal transformation in two
dimensions
consists essentially in a redefinition of the fields and therefore
the physical content of the theory should not depend on it.
In particular, if one defines a rescaled metric
$$\gmn=\ef\ghmn,\eqno(2)$$
the action (1) becomes
$$S=\dx\left[\ef R+4\l^2-\ha\sum(\nabla f_i)^2\right].\eqno(3)$$
However, the action (3) admits only flat solutions (in Rindler \coo)
with
non-trivial dilaton. It has therefore been argued that the theory
defined by (1) should be
trivial, since it is equivalent to flat space under conformal \tran
and in
particular does not admit \hr, since the conformal anomaly is of
course zero
for flat space [3], at variance with the results of [1]. In this
letter, we
wish to clarify this point, by observing that in order to solve the
puzzle,
one has to take into due account
the role of the dilaton field, which should be considered on the same
footing as the metric in the discussion of the structure of  the
spacetime.
In particular, the requirement of reality of the dilaton field
implies
that
even flat solutions have a non-trivial structure, since one is forced
to
impose a boundary to the spacetime on the curve where $\ef$ changes its
sign (from a four-dimensional point of view the line $e^{-2\phi}=0$
corresponds to the origin of the radial coordinate $r$).
This fact prevents the possibility of performing a change of \coo to
put the
Rindler metric in a \min form. But it  is well-known that to Rindler
\coo
is associated a constant flux of \hr, which is in this case of purely
topological origin, and whose value turns out to coincide with the
one
calculated from (1). This mechanism is very similar to that discussed
in [4]
for two-dimensional \ads spacetime, in the context of the
Jackiw-Teitelboim
model.

{}From our discussion will also  emerge that the CGHS vacuum is
semiclassically
unstable, unless one imposes a priori a cosmic censorship hypothesis
(which
can however be justified from a four-dimensional point of view). In
doing
that,
we clarify some results obtained in [5] by means of a moving mirror
model.
Our arguments will also permit us to consistently define a mass for
the
dilaton Rindler spacetimes, which coincides with that of the
conformally
related CGHS solution.

The field equations stemming from (3) are:
$$\eqalign{&R=0,\qquad \nabla^2 f_i=0,\cr
&(\gmn\nabla^2-\nabla_\m\nabla_\n)\ef=2\l^2 +T_{\mu\nu}^{(f)},
\cr}\eqno(4)$$
where $T_{\mu\nu}^{(f)}$ is the energy-momentum tensor for the
fields $f_i$.
The general static solutions of these equations in the \sch gauge are
given, for vanishing $f_i$, by a
locally flat metric and a non-trivial dilaton, namely:
$$\eqalign{\ds  -(2\l r&-c)dt^2+(2\l r-c)^\mo dr^2,\cr
&\ef=2\l r+d.\cr}\eqno(5)$$
Without loss of generality, one can take $d=0$. It is important to
notice that
the static solutions arise naturally in Rindler \coo [6]. A change of
\coo
$\s=\l^\mo\sqrt{2\l r-c}\ \cosh\l t$, $\t=\l^\mo\sqrt{2\l r-c}\
\sinh\l t$
brings the metric to the \min form $\ds -d\t^2+d\s^2$, but the
dilaton
becomes
time-dependent, $\ef=c+\l^2(\s^2-\t^2)$.

The central point of this paper is however the observation that, if
one wishes
to have a real dilaton, one is forced to cut the spacetime at the
curve where
$\ef$ changes its sign ($r=0$ in the \coo (5)), so that one cannot
obtain
the
full \min space even by changing \coo.
One can interpret the curve where $\ef$ vanishes as a
singularity of
spacetime. Indeed this  curve corresponds to a true curvature singularity
in the CGHS model, as one can easily verify performing the rescaling
(2).
For $c>0$, this singularity is shielded by a horizon at $r=c/2\l$ and
is
spacelike, while for  $c\le 0$ it is naked and timelike. In the
case $c=0$
it coincides with the coordinate singularity and is lightlike.

It is also possible to assign a mass to the solutions (5) by means of
the ADM
procedure. In fact, one can define a conserved mass function [7]:
$$M={1\over 4\l}\left[4\l^2\ef-(\nabla\ef)^2\right],$$
whose value at infinity gives the ADM mass of the solution:
$$M=\l c.$$
Of course, the nonvanishing value of the mass is due to the
contribution of
the dilaton, which cannot be separated from that of the metric.

The temperature at the horizon of the metric (5)
can be easily obtained and is independent of
$c$. Its value is given  by
$$T={\l\over 2\p}.\eqno(6)$$
Both values of $M$ and $T$
agree with those of the conformally related CGHS solutions.

The solution (5) can thus be interpreted as a \bh of mass $M$ and
temperature
$T$, with a singularity at the origin of \coo. If $M$ is negative, a
naked
singularity is present. The state with $c=0$ may be interpreted as
the
\gs of
the theory. From a two-dimensional point of view, this choice looks
quite
arbitrary, since the temperature is independent of the mass. A
justification
for this choice is provided if one considers the dilaton \sing as a
real
\sing and imposes a sort of cosmic censorship conjecture, which rules
out
the $M<0$ solutions. Indeed, the negative mass solutions correspond
to
naked
singularities of the original four-dimensional theory.

The black hole solutions (5) have been first derived by Mann, Shiekh
and Tarasov in the context of a $R=T$ two-dimensional gravity theory
in ref.[8], where
a pointlike mass with $T\propto M\delta(r)$ acts as source for the
gravitational
field. Our derivation is however different, in fact from Eqs. (4)
one easily realizes that the coupling of matter to the gravitational
degrees of freedom is encoded not in the equation for the Ricci
scalar but in the equation for the dilaton. This is of course related
to
the central role played by the dilaton in our model.

The results obtained till now can be readily translated into a
conformal
gauge, which is more apt to the discussion of the \hr and to the
comparison
with the CGHS model.
In light-cone \coo, the metric has the Rindler form:
$$\ds-\ely dy^+dy^-,\eqno(7)$$
with dilaton
$$\ef=c+\ely.\eqno(8)$$
The \coo $y^\pm$ are related to the \sch\coo by
$$y^\pm=t\pm{1\over 2\l}\ln(2\l r-c).$$
Notice that   the boundary at $r=0$ is not
visible in these \coo if $c>0$, since they cover only the region
of the \bh\st outside the horizon at $r=c/2\lambda$.

Once the interpretation of solution (5) as a black hole has been
established,  at the semiclassical level one would naively expect
this black hole to evaporate. The emergence  of the Hawking
radiation
in our simple two-dimensional gravity model is however a point which
deserves a careful study. Indeed conformal anomaly arguments have been
used
to argue that, being the space everywhere flat, there is no Hawking
radiation in this model [3].
In the following we will demonstrate, using standard quantization
techniques, that the semiclassical dynamics  of the black holes
naturally gives rise to particle creation with thermal spectrum.
Even though this radiation cannot be interpreted in the usual way
as the Hawking radiation seen by  an observer in  an asymptotically
flat region, its existence sheds some light on the properties of
the CGHS model and is crucial to understand the physical equivalence
of the models related by the rescaling (2).

Let us assume that the black hole is formed by the collapse of a
$f$-wave. For example one can easily construct, using light-cone
coordinates, solutions describing the collapse of a $f$-shock wave
at $y^+=y^{+}_0$: for $y^+ \le y^+_0$ the solution is given by
the vacuum solution ($c=M=0$), whereas for $y^+\ge y^{+}_0$ a black
hole is formed.
The key point is that for the vacuum solution  the boundary at
$r=0$ is lightlike so that one can introduce coordinates
$$x^\pm=\pm {1\over \lambda}e^{\pm \lambda y^\pm}, \eqno (9)$$
to put the solution in the \min form:
$$ds^2= -dx^+dx^-, \qquad e^{-2\phi}= -\lambda^2 x^+x^-.\eqno(10)$$
The spacetime is of course not complete and has to be thought as
ordinary \min space endowed with a lightlike boundary.
For $y^+ \ge y^+_0$ a black hole forms and the boundary at $r=0$
becomes spacelike. The presence of a spacelike boundary prevents the
possibility of performing the change of \coo (9) to put  the
metric in the \min form. The latter  has therefore to be given in
the form (7).
The coordinates $y^\pm$ cover  only the region outside
the event horizon which in terms of the coordinates $x^\pm$ is
described  by the lines $x^+=0,x^-=0$.
The \coo $y^\pm$ give the Rindler coordinatization of \min space.
The lines of constant $y=(y^+-y^-)/2$ represent the world lines of
uniformly accelerated observers in \min space, with proper
acceleration given by $\lambda e^{-\lambda y}.$

Let us now consider the quantization of a single scalar field $f$ in
the
fixed background defined by (7). This corresponds to perform the
quantization of a scalar field in Rindler space which is by now
well known and can be found in the standard textbooks [9].
The quantum field $f$ can be decomposed both in terms of positive
frequency \min and Rindler modes. The natural timelike
coordinates
of the two spaces are related in such a way that a field mode which has
positive frequency according to a \min observer inevitably
becomes a mixture of positive and negative frequencies according
to a Rindler observer. This mixing is then interpreted as particle
creation. The \min vacuum $|0_M>$ will therefore appear to
a Rindler observer as filled with thermal radiation.
The detailed structure of the Hawking radiation is encoded in the
Bogoliubov transformation relating $|0_M>$ to the Rindler vacuum
$|0_R>$. The final outcome is that a Rindler observer will perceive
the \min vacuum as thermally populated with spectrum:
$${1\over{e^{2\pi\omega/\lambda}-1}}.\eqno(11)$$
This is precisely the Planck spectrum for radiation at temperature
given by (6).
When integrated it gives the total flux of $f$-particle
energy:
$$G={\lambda ^2\over 48},\eqno(12)$$
which is in agreement with the CGHS result.
Our results simply state that the semiclassical behavior of our
black holes is described by a quantum field theory of accelerated
observers in \min space. The physical interpretation of the associated thermal
radiation deserves however careful analysis.
It is in fact evident that it cannot be interpreted in the usual
way, as  the Hawking radiation seen by an observer in an
asymptotically flat spacetime region. This prevents the usual
interpretation of the Hawking effect as the distortion, due to
the black hole geometry, of incoming
quantum modes defined in a flat region into outcoming ones
defined in another flat region.
The point is that in our model the black hole geometry is truly
equivalent to  that seen by accelerated observers in \min
space. The event horizon of the black hole has to be considered as
an acceleration horizon.
This discussion clarifies a point which appears rather puzzling
in the CGHS model. The CGHS black holes have temperature and Hawking
flux, given by (6) and
(12) respectively, independent of the mass of the black hole;
 there is no explanation
of this fact in CGHS theory. In our model a natural explanation
of this fact is at hand: the thermal properties are independent of
the
mass because the semiclassical dynamics of the black hole is
equivalent to that of accelerated observers, they depend only on
the parameter $\lambda $ which defines the
proper acceleration of these observers.
Moreover our results restore some of the physical intuition about the
equivalence of the models under the rescaling of the metric (2).
What we have found is that the models (1) and (3) have an {\rm
equivalent}
classical and semiclassical dynamics. Indeed the two metrics, being
related by a Weyl rescaling, have the same causal structure (the
same Penrose diagram) and at the semiclassical level have the
same thermal properties.

At first glance our result seems to contradict the well-known
relationship
between conformal anomaly and Hawking radiation. Being the black hole
spacetime everywhere flat the conformal anomaly vanishes and there
should be no Hawking radiation.
However the presence of thermal radiation in our model is related
to a topological effect which is independent of the presence of
a conformal anomaly. On the other hand the possibility of having
Hawking radiation in globally flat spaces is related to the
anomalous transformation law of the quantum  energy-momentum tensor
$T_{--}$ under coordinates change.

To conclude this letter let us  comment about the question
of the stability of the ground state in our model and in the
related CGHS model. Soon after the discovery of the CGHS model
it was realised that the semiclassical ground state of this model
is unstable and plagued by the presence of naked singularities [2].
The meaning and the origin of this instability has been further
clarified
in Ref. [5], where a dynamical moving mirror in flat \st
was used to model
the semiclassical evolution of a CGHS black hole. It was found that
vacuum solution do exist which describe a forever accelerating
mirror resulting in an unphysical process of a forever
radiating black hole.
These features have a natural explanation in the context of our model.
Differently from the CGHS case, for which the classical vacuum
(the so called linear dilaton vacuum) is a perfectly regular and
geodesically complete spacetime, in our model one can define
consistently a classical ground state only using a cosmic censorship
conjecture to rule out the states with negative mass.
Moreover this  vacuum does not describe a complete spacetime
but Minkowsky space with a light-like boundary.
The presence of this boundary is a potential source of instability of
the vacuum. This can be easily understood if one allows in the
spectrum the states with negative mass (thus giving up the cosmic
censorship conjecture). The dynamics can be now described in terms of
the evolution of a timelike boundary in Minkowski space, it is therefore
equivalent to the dynamical moving mirrors  discussed in ref.[5]. There the
authors considered a boundary
on the curve where $\ef=C$, where $C$ is an arbitrary positive
constant. Comparing with the expression (10) one easily sees that
this is equivalent to consider our solutions with negative mass given
by $M=-C\lambda$ with a boundary on the curve where
$\ef=0$. The runaway solutions describing a forever accelerating
moving mirror found in [5]
are therefore related to the infinite tower of black hole states
with negative mass, i.e to spacetimes with a timelike boundary at
$\ef=0$.
Thus, unless one uses a cosmic censorship hypothesis to discard
these states as unphysical, one  has to deal with processes in which
the black hole mass becomes arbitrarily negative.
A similar point of view, stating the role of spacetime singularities
acting as regulators forbidding  the appearance of unphysical states
in the spectrum, has been put forward recently by Horowitz and Myers [10].

\smallskip
\beginref

\ref [1] C.G. Callan, S.B. Giddings, J.A. Harvey and A. Strominger,
\PRD {\bf 45}, 1005 (1992);

\ref [2] J.G. Russo, L. Susskind and L. Thorlacius \PRD {\bf 46},
3445 (1992); \PRD{\bf 47}, 533 (1993).

\ref [3] D. Cangemi and R. Jackiw, \PRD{\bf 50}, 3913 (1994);
T. Fujiwara, Y. Igarashi and J. Kubo, \PLB {\bf 316}, 66 (1993).

\ref [4] M. Cadoni and S. Mignemi, \PRD  (in press), hep-th 9410041.

\ref [5] T.D. Chung and H. Verlinde, \NPB {\bf 418}, 305 (1994).

\ref [6] W. Rindler, {\it Essential Relativity} (Springer-Verlag, 1969).

\ref [7] R.B. Mann, \PRD {\bf 47}, 4438 (1993);
S. Mignemi, preprint INFNCA-TH-94-28.

\ref [8] R.B. Mann, A. Shiekh and I. Tarasov, \NPB {\bf 341}, 134 (1990);
R.B. Mann, Gen. Rel. Grav. {\bf 24}, 433 (1992).

\ref [9] N.D. Birrell and P.C.W. Davies, {\sl Quantum fields in
curved space} (Cambridge Un. Press, 1982).

\ref [10] G.T. Horowitz and R.C. Myers, gr-qc 9503062.

\endref
\end